\documentclass[paper]{ite}                  % for Paper
\usepackage{graphicx}
\usepackage[usenames]{color}
\usepackage{times}
\usepackage{epsfig}
\usepackage{graphicx}
\usepackage{amsmath}
\usepackage{amssymb}
\usepackage{wasysym}
\usepackage{tweaklist}
\usepackage{array}
\usepackage{verbatim}
\usepackage{enumitem}
\usepackage{multirow}
\usepackage{microtype}
\usepackage[]{algorithm2e}
\usepackage{algpseudocode}
\usepackage[bottom]{footmisc}

\usepackage{lipsum}

\newcommand\blfootnote[1]{%
  \begingroup
  \renewcommand\thefootnote{}\footnote{#1}%
  \addtocounter{footnote}{-1}%
  \endgroup
}

\setlist{nolistsep}

\usepackage{array}
\newcolumntype{L}[1]{>{\raggedright\let\newline\\\arraybackslash\hspace{0pt}}m{#1}}
\newcolumntype{C}[1]{>{\centering\let\newline\\\arraybackslash\hspace{0pt}}m{#1}}
\newcolumntype{R}[1]{>{\raggedleft\let\newline\\\arraybackslash\hspace{0pt}}m{#1}}

\title{Strategies for Searching Video Content with Text Queries or Video Examples}
\subtitle{Features, Semantic Detectors, Fusion, Efficient Search and Reranking}
\authorlist{%
 \authorentry{Shoou-I Yu*}{s}{Company1} 
 \authorentry{Yi Yang}{m}{Company2} 
 \authorentry{Zhongwen Xu}{s}{Company2}
 \authorentry{Shicheng Xu}{s}{Company1}
 \authorentry{Deyu Meng}{m}{Company3}
 \authorentry{Zexi Mao}{m}{Company1}
 \authorentry{Zhigang Ma}{m}{Company1}
 \authorentry{Ming Lin}{m}{Company1}
 \authorentry{Xuanchong Li}{s}{Company1}
 \authorentry{Huan Li}{m}{Company1} 
 \authorentry{Zhenzhong Lan}{s}{Company1}
 \authorentry{Lu Jiang}{s}{Company1} 
 \authorentry{Alexander G. Hauptmann}{m}{Company1}
 \authorentry{Chuang Gan}{s}{Company4}
 \authorentry{Xingzhong Du}{s}{Company5}
 \authorentry{Xiaojun Chang}{s}{Company2}
}
\affiliate[Company1]
 {Language Technologies Institute, Carnegie Mellon University.}
 {Pittsburgh, PA, USA.}
\affiliate[Company2]
 {Centre for Quantum Computation and Intelligent Systems, University of Technology Sydney.}
 {Sydney, NSW, Australia.}
 \affiliate[Company3]
 {School of Mathematics and Statistics and Ministry of Education Key Lab of Intelligent Networks and Network Security, Xi’an Jiaotong University.}
 {Xi'an, China.}
\affiliate[Company4]
 {Institute for Interdisciplinary Information Sciences, Tsinghua University}
 {Beijing, China.}
\affiliate[Company5]
 {School of Information Technology and Electrical Engineering, The University of Queensland.}
 {Brisbane, QLD, Australia.}

\acceptancePreIN{} % Do not modify this line.
\videoFlag{0} % If any videos are included, select "1". Otherwise, select "0".
\begin{document}
\begin{abstract}
The large number of user-generated videos uploaded on to the Internet everyday has led to many commercial
video search engines, which mainly rely on text metadata for search.
However, metadata is often lacking for user-generated videos, 
thus these videos are unsearchable by current search engines.
Therefore, content-based video retrieval (CBVR) tackles this metadata-scarcity problem by directly analyzing the visual and audio streams of each video.
CBVR encompasses multiple research topics,
including low-level feature design, feature fusion, semantic detector training and video search/reranking.
We present novel strategies in these topics to enhance CBVR in both accuracy and speed under different query inputs, including pure textual queries and query by video examples.
Our proposed strategies have been incorporated into our submission for the TRECVID 2014 Multimedia Event Detection evaluation,
where our system outperformed other submissions in both text queries and video example queries,
thus demonstrating the effectiveness of our proposed approaches.
\end{abstract}
\maketitle

\blfootnote{*$\,\,$Authors are sorted in reverse alphabetical order.}
\section{Introduction}
As we see an unprecedented growth of user-generated videos on the Internet, it is crucial to have an effective 
indexing and searching mechanism for these videos.
To perform search, current existing video search engines mainly rely on user-generated text metadata.
However, text metadata is often not a comprehensive representation of the video as:
1) users often do not provide metadata, and 
2) even if users do provide metadata, a user cannot possibly annotate all facets of the video.
Therefore, content-based video retrieval (CBVR), which directly analyzes the visual and audio channels of a video to perform search, has attracted the attention of many researchers
and the annual TRECVID Multimedia Event Detection (MED) evaluation \cite{TRECVID14} was created.
In this independent evaluation, participants design systems
which utilize the wealth of information in the visual and audio channels
to perform effective and efficient content-based video search for different query types, including 
1) text queries and
2) query by video example.

Compared with the already mature text-based search, 
CBVR is significantly more challenging.
One big challenge is the \emph{low-level feature extraction} challenge.
A big problem with raw visual and audio channels is that 
videos which depict similar semantics will still look very different if one directly compared the raw values of the two channels.
To make matters worse, user-generated videos are usually very unstructured, have low resolution, severe camera motion, and very large variability.
Therefore, representing videos with features which have certain invariance and generalization capabilities is a crucial part of CBVR. 
Another challenge is the \emph{text/video semantic gap} challenge~\cite{jiang2015bridging}.
The main problem is that the aforementioned feature representations for video often do not contain semantic information, 
but to query video data with textual queries, it is crucial to bridge the semantic gap between a pure text-based query and the non-semantic representation of a video.
The final challenge is the \emph{indexing/search} challenge.
As new features are used to represent the visual and audio channels, traditional text-search techniques are not directly applicable,
and new search techniques need to be developed. To enable search over large video collections, these new search techniques have to be both effective and efficient.

In light of the aforementioned challenges, we propose multiple strategies to tackle these problems.
For the \emph{low-level feature extraction} challenge, we propose two different features to significantly enhance CBVR performance.
The first feature is a variant of the Improved Dense Trajectory feature \cite{imtraj,Lan141} (Section~\ref{sec:imtraj}), 
and the second feature is a deep learning feature (Section~\ref{sec:koala}) trained on ImageNet \cite{Den09,Kri12} data.
For the \emph{text/video semantic gap} challenge, we propose a method which utilizes large amounts of weakly-labeled videos
to learn semantic concept detectors encompassing a large vocabulary (Section~\ref{sec:concepts}).
This enlarged vocabulary is crucial in bridging the semantic gap between a text-query and non-semantic video representations.
For the \emph{indexing/search} challenge,
we first propose to utilize Explicit Feature Maps \cite{And12} and Product Quantization \cite{Jeg11} to perform efficient yet effective
video search (Section~\ref{sec:pq}).
We then propose a novel fusion method called Multistage Hybrid Late Fusion (MHLF) to effectively fuse search results from multiple feature modalities (Section~\ref{sec:ming}). 
Finally, we propose a self-paced reranking method \cite{LuJ141} to automatically enhance search results through pseudo-relevance feedback (Section~\ref{sec:reranking}). 
The aforementioned methods were all integrated into our TRECVID MED 2014 system,
which was the leading system in all eight MED subtasks, thus demonstrating the effectiveness of our proposed strategies.
%As shown in Figure~\ref{fig:prespecified} and Figure~\ref{fig:adhoc}, our system is the best in all four exemplar conditions,
%which clearly shows the effectiveness of our methods.

In the following sections, we first give an overview of a general CBVR system and related work in Section~\ref{sec:related}.
Then we summarize our results in the TRECVID MED 2014 task in Section~\ref{sec:medtask}.
Details of each proposed strategy are given in Sections~\ref{sec:feat}, \ref{sec:gap} and \ref{sec:indexing}.
Finally, Section~\ref{sec:conclusion} concludes the paper.

\section{Content-based Video Retrieval Preliminaries}
\label{sec:related}

\begin{figure}[tp]
\centering
\includegraphics[scale=0.35]{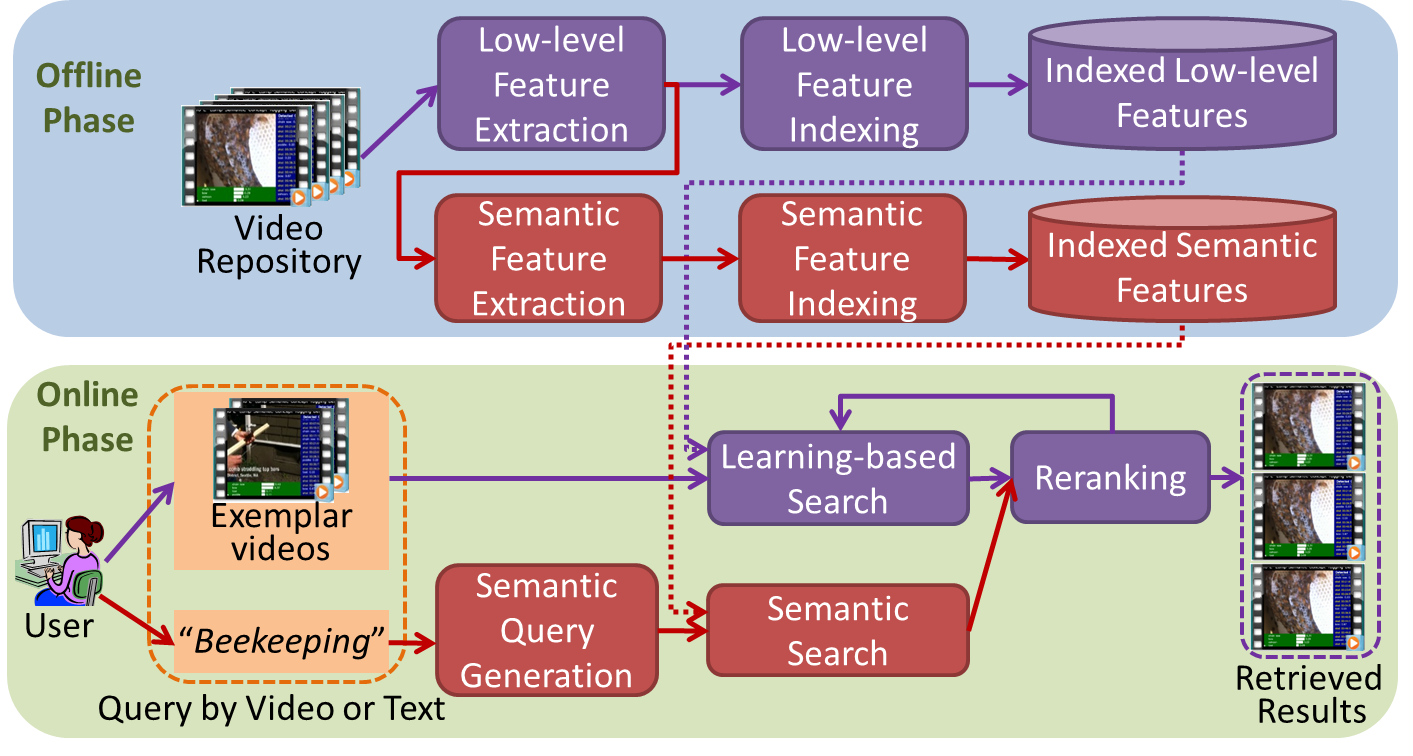}
\caption{Pipeline of a general CBVR system. The purple boxes and arrows correspond to components designed for querying by video examples.
The red boxes and arrows corresponds to components designed for text queries.}
\label{fig:pipeline}
\end{figure}

A general pipeline of a CBVR system is shown in Figure~\ref{fig:pipeline}.
There are mainly two phases: the offline phase and the online phase.
In the offline phase, low-level and semantic features are extracted for a large video repository and indexed so that the online phase is sufficiently efficient.
The semantic features are predictions of semantic concept detectors, which takes low-level features
as input and predicts whether a given concept such as dog, cat, or car exists in a video.
In the online phase, users will provide different types of queries to search for relevant videos.
There are mainly two types:
\begin{enumerate}
\item Query by video example: The user provides one or multiple example videos to search for related videos.
\item Text queries: The user types in a pure text query to search for videos of interest. 
\end{enumerate}
As the input from different query types are of different modalities (i.e. videos or text), different kinds of features and search techniques are
designed for each case.
For the query by video examples case,
the learning-based search component retrieves related videos by first
training a model which distinguishes the exemplar videos from the non-related videos.
The model is trained based on the features extracted in the offline phase.
Then, the model is applied to the video repository to search for other related videos.
%For convenience we assume that the user-provided exemplars are in the video repository, so the learning-based search component 
%can directly take the features for the examples from the index.
%Based on these features, a model is learned to distinguish related videos from non-related videos.
Lastly, the search results goes through an iterative reranking process, which performs pseudo-relevance feedback to automatically improve the search results.
For searching by text queries, the first step is semantic query generation,
where the text query is mapped to the system vocabulary.
The system vocabulary constitutes of all concepts that could be detected by the available semantic concept detectors.
Then the generated semantic query is utilized to perform semantic search.
The initial ranked list also goes through the reranking process to acquire a more accurate ranked list.
In the following sections, we will explain the details and also briefly review the related work for each component.

\subsection{Searching by Video Examples}
For the query by video examples scenario, low-level features combined with discriminatively learned search models
play the key role in achieving good performance \cite{natarajan2012multimodal,informedia14}.
In the following sections, we will review the related work on these two topics.

\subsubsection{Low-level Features}
Many low-level audio and visual features have been utilized to enhance CBVR performance.
The most popular low-level audio feature used is the Mel-Frequency Cepstral Coefficients (MFCC) \cite{informedia14},
which have been shown to be the most cost-effective feature \cite{Lan14}.
Other audio features including Acoustic Unit Descriptors \cite{Cha11} and Large-scale Pooling Features \cite{metze2014improved} have also been utilized.

Low-level visual features can be split into two categories: \textit{static image features} and \textit{motion features}.
Static image features are essentially image-based features extracted from all or selected frames of a video.
The temporal relation between the frames are not taken into account.
Before the introduction and success of deep features, mainstream static image features were frequently hand-crafted SIFT-based features \cite{SIFT,CSIFT}.
Currently, deep convolutional neural network (DCNN) features \cite{informedia14,Zho14} are significantly outperforming hand-crafted features and represent the current mainstream.
In this paper, we present architectural improvements for static-image deep-networks to enhance CBVR performance.

Motion features utilize the temporal relations between frames to capture motion characteristics of a video.
Optical flow is typically used to compute motion features.
Currently, one of the most effective features is Improved Dense Trajectories (IDT) \cite{Hen13}, which significantly outperforms the previously proposed
popular motion features such as Space Time Interest Points (STIP) \cite{STIP} and Motion SIFT (MoSIFT) \cite{che09}.
In this paper, we present two enhancements which further improves the performance of IDT.

One problem with the previously mentioned low-level features 
is that they will generate a different number of feature vectors depending on the length, resolution and contents of the video,
thus leading to varying length vector representations for each video.
It is very difficult to compare two videos with different length representations.
Therefore, the varying length representations of each video need to be converted to a fixed-length vector representation, thus
many different encoding/pooling techniques have been proposed, including Bag-of-Words (BoW) \cite{sivic2003video} and Spatial Pyramid BoW (SpBoW) \cite{Laz06}, 
Fisher Vectors (FV) \cite{KCh11,fisher}, and Vector of Locally Aggregated Descriptors (VLAD) \cite{VLAD}.
FV and VLAD are the current mainstream encoding methods \cite{Zho14}.

Overall, a ``complete'' feature is a combination of 
a low-level feature and an encoding method. For example, SIFT can be encoded
with SIFT-SpBoW or SIFT-FV, and MFCC can also be encoded with BoW (MFCC-BoW) or FV (MFCC-FV) respectively.
Once these encodings have been computed, they are indexed for the subsequent learning-based search.

\subsubsection{Learning-based Search}
There are two key components to learning-based search: the learning component, and the fusion component.

Utilizing machine learning models such as Support Vector Machines (SVM) \cite{natarajan2012multimodal,ChangYXY15a} and Kernel Ridge Regression (KRR) \cite{informedia14}
has shown to be very effective for querying with video examples.
The main idea is to treat the example videos as positive training data, and when
combined with a large pool of negative videos, a classifier can be trained to determine whether an input testing video is relevant or not.

Fusion enables the incorporation of search results from different features which capture the multiple aspects of a video.
The main challenge of fusion is to effectively estimate the reliability of each feature source so that the fusion algorithm
knows which features to rely more on when dealing with different videos. 
Many fusion method such as early fusion, late fusion, double fusion \cite{Lan13} and 
other more complex methods \cite{ye2012robust} have been proposed.
%For early fusion, features from different modalities are first concatenated into one long vector before training a single classifier.
%For late fusion, a classifier is trained for each feature. In the search phase, prediction results are first computed for each feature.
%Then, the fusion algorithm will perform a weighted sum over the different prediction results.
%Double fusion simply combines early fusion and late fusion results.
In this paper, we propose a Multistage Hybrid Late Fusion method, which shows superior performance and robustness over other fusion methods.

\subsubsection{Reranking}
Reranking utilizes pseudo-relevance feedback (PRF) to automatically enhance an initial rank list.
The intuition of PRF is that the top-ranked results in an initial rank list are highly likely to be correct,
and adding these instances back into the training set may improve performance.
This simple method has shown to be effective in many different scenarios.
However, previous PRF methods usually operate on a single ranked list \cite{joachims1996probabilistic},
but the CBVR task inherently outputs multiple ranked lists from different features,
and effectively fusing these ranked lists becomes a challenging task \cite{MMPRF}.

In this paper, we introduce self-paced reranking, which further improves the performance of existing PRF approaches.
Our system incorporates MMPRF \cite{MMPRF} and SPaR \cite{LuJ141} to conduct reranking, in which MMPRF is used to assign the starting values, and SPaR is used as the core reranking algorithm. The reranking is inspired by the self-paced learning proposed by Jiang et al.\cite{LuJ141}, in that the model is trained iteratively as opposed to simultaneously. Our methods are able to leverage high-level and low-level features which generally leads to increased performance \cite{LuJ12}. The high-level features used are ASR, OCR, and semantic visual concepts. The low-level features include DCNN, IDT and MFCC features. %We did not run PRF for SQ and 100Ex runs. For SQ run it is because our SQ run is essentially the same as our 0Ex run. For 100Ex it is because the improvement on the validation set is less significant.

\subsection{Searching with Text Queries}
This scenario takes a pure-text query as input, and outputs a ranked list of relevant videos. 
It is an interesting task because it resembles a real-world video search scenario, where users typically search videos by using query words instead of providing example videos. 

The main challenge of the text-to-video search scenario is to bridge the semantic gap between text and video.
In current state-of-the-art systems, this gap is usually bridged with automatic speech recognition (ASR), optical character recognition (OCR),
and semantic concept detectors.
Semantic concept detectors are trained to detect whether a certain object, scene, or action exists in a video or not.
Given a pool of concept detectors, these detectors can be applied on an input video
to acquire a semantic feature representation of the video, which corresponds to the confidence score
of detecting a concept in the video.
This feature representation is very different from the low-level feature representations,
where each dimension in the vector does not have a clear semantic meaning.
Popular datasets to train concept detectors include the ImageNET \cite{Den09} dataset, the SUN397 \cite{xiao2014sun} scene dataset,
and the TRECVID Semantic Indexing (SIN) \cite{TRECVID14} dataset.
To train effective static-image-based detectors, the current mainstream is deep convolutional neural network models \cite{Kri12,ren2015faster}.
To train video-based detectors, the current mainstream is combining deep static-image detectors with motion features such as Improved Dense Trajectories \cite{Hen13}.

\begin{comment}
\begin{figure}[tp]
\begin{center}
\begin{tabular}{ccc}
\includegraphics[width=8.0cm,angle=0]{ek0.png} \\
\end{tabular}
\end{center}
\caption{
The framework of the text query system\cite{MMPRF}.
}
\label{fig:ek0}
\end{figure}
\end{comment}

According to Jiang et al.\cite{MMPRF,jiang2015bridging}, a text-to-video search system consists of three major components, 
namely Semantic Query Generation (SQG), Semantic Search and 
Reranking/PRF as shown in Figure~\ref{fig:pipeline}.
The Semantic Query Generation component translates the description of the user's information need into a set of multimodal system queries that can be processed by the system. There are two challenges in this step. Since the semantic vocabulary of the system is usually limited, the first challenge is to map the user's query words into the system vocabulary. The second challenge is assigning a given query word its modality as well as its weight associated with that modality. A preliminary study of these challenges is detailed in Jiang et al.~\cite{jiang2015bridging}.

The semantic search component retrieves multiple ranked lists for a given text query. Our system incorporates various retrieval methods such as the Vector Space Model, tf-idf, BM25, language model \cite{CZh01}, etc. Surprisingly, a better retrieval model on worse features actually outperforms a worse retrieval model on better features. This observation suggests that the role of retrieval models in our semantic search system may be underestimated in much current research. After retrieving the ranked lists for all modalities, we apply a normalized fusion to fuse different ranked lists according to the weights specified in SQG.

Reranking is also performed for text query search.
One key advantage of reranking is that it ``bridges'' the semantic search and the learning-based search \cite{MMPRF}.
Once the text query search component generates an initial ranked list, the positives in this ranked list
can be used to perform learning-based search, which can often further improve search performance.

\section{TRECVID Multimedia Event Detection 2014}

\label{sec:medtask}
\begin{figure}[tp]
\begin{center}
\begin{tabular}{ccc}
\includegraphics[width=8.0cm,angle=0]{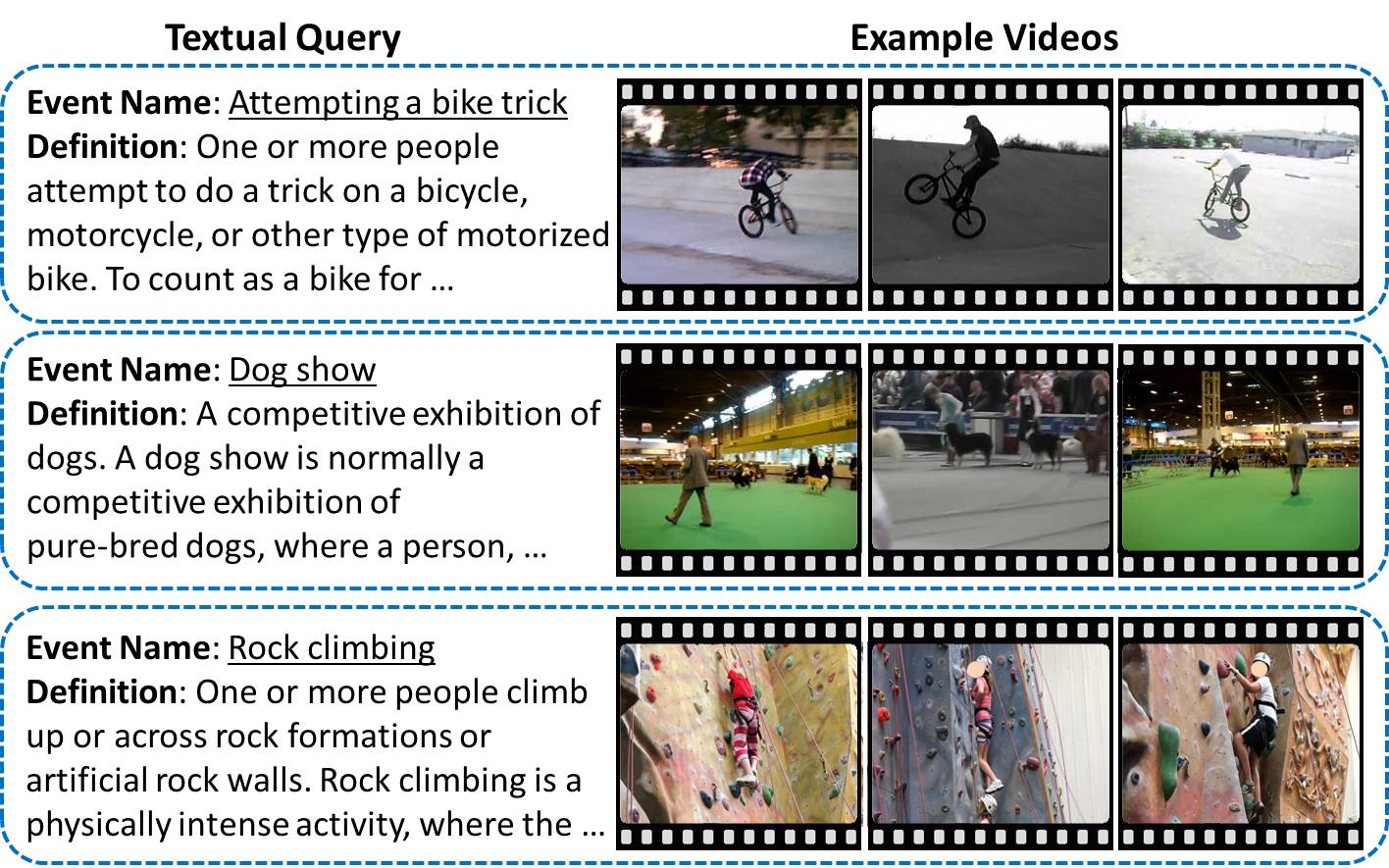} \\
\end{tabular}
\end{center}
\caption{
Examples queries from the MED14 task. Each query consists of a textual description and video examples.
}
\label{fig:medexamples}
\end{figure}

The TREC Video Multimedia Event Detection (MED) \cite{TRECVID14} task is a standardized task held every year since 2010
to evaluate the performance of different CBVR systems on the MED task.
Different CBVR systems are presented with multiple queries, and the CBVR system needs to retrieve relevant videos from the evaluation set.
Example queries are shown in Figure~\ref{fig:medexamples}.
The query consists of two parts, the textual query and the video examples.
Different parts of the query will be utilized for different query settings as described below.
%For a fair comparison, all competitors follow the standard data split specified by the organizers.
For the TRECVID MED14 task, the organizers split the data set into four
standard sets: the positive examples, the background set, the validation set and the testing set.
The positive examples are videos
relevant to a queried event.
The number of positive examples will vary according to the four different query settings defined by the organizers: 
\begin{enumerate}
\item Semantic Query (SQ): Only the textual query was given.
\item 0 Exemplar (000Ex): The textual query and background videos were given. No positive videos were given.
\item 10 Exemplar (010Ex): In addition to what was given for 000Ex, 10 positive videos were given.
\item 100 Exemplar (100Ex): In addition to what was given for 000Ex, 100 positive videos were given.
\end{enumerate}
The background set, which can be viewed as the negative videos, contained 4992 videos.
The validation set, which is also know as MEDTEST14, contained around 24,000 videos.
The testing set contained around 198,000 videos (8000 hours of video) which does not contain any text metadata.
In the competition, competitors trained their system on the positive and background sets, and then tuned the system on the validation set.
Finally, the resulting system performed search over the testing set and the results were submitted to the organizers.
Label information was only available for the positive, background and validation sets.
The labels for the testing set were never released to prevent overfitting on the testing set.
In the competition, there were two types of queries, pre-specified and ad-hoc.
For pre-specified queries, the names of events were given a few months beforehand,
so participants could design specialized detectors for these events.
On the other hand, ad-hoc queries were given a few days before the deadline,
leaving no time to design specialized detectors.
There were a total of 20 events/queries for the MED14 pre-specified run
and 10 events for the MED14 ad-hoc run.
%Queries include a wide variety of events, such as ``Birthday Party'', ``Repairing an Appliance'', ``Dog Show'', and ``Beekeeping''.
The evaluation metric used was Mean Average Precision (MAP) \cite{TRECVID14}.

\begin{figure}[tp]
\begin{center}
\begin{tabular}{ccc}
\includegraphics[width=8.0cm,angle=0]{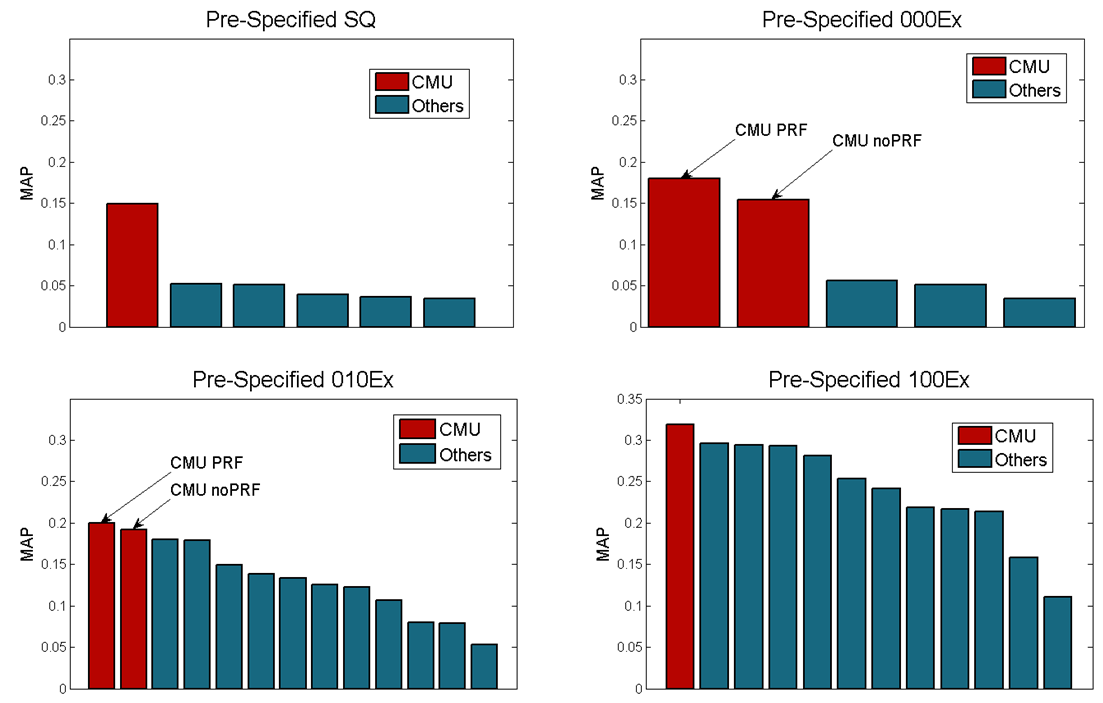} \\
\end{tabular}
\end{center}
\caption{
Official\cite{TRECVID14} MAP performance on the MED14 testing set in different settings for pre-specified events.
}
\label{fig:prespecified}
\end{figure}
\begin{figure}[tp]
\begin{center}
\begin{tabular}{ccc}
\includegraphics[width=8.0cm,angle=0]{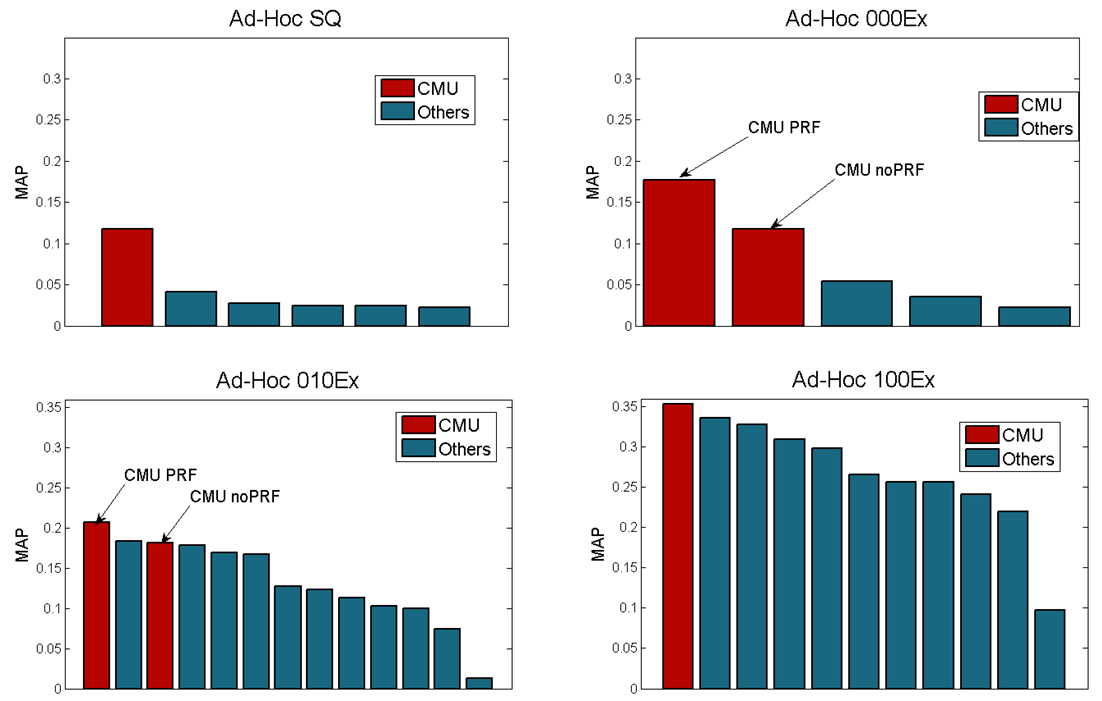} \\
\end{tabular}
\end{center}
\caption{
Official\cite{TRECVID14} MAP performance on the MED14 testing set in different settings for ad-hoc events.
}
\label{fig:adhoc}
\end{figure}

\begin{table}
\centering
\begin{tabular}{|c c c|}
\hline
\hline 
 Metric: MAP &   100Ex & 010Ex \\\hline
  IDT & 0.274 & 0.133  \\ 
  IDT + SPM   & 0.286  & 0.136\\
  MIFS (L=0,2,5) & 0.297&  0.153  \\
  MIFS (L=0,2,5) + STED  & \textbf{0.298} &  \textbf{0.162}  \\
\hline
\end{tabular}
\caption{\label{tab:mifsmed}Performance comparison of MIFS and STED over baseline methods.}
\end{table}

\subsection*{CMU MED14 Submission Overview}

For MED14, we had a system\cite{informedia14} for text queries (SQ, 000Ex) and another system for query by video examples (010Ex, 100Ex).
For text queries, our system utilized ASR, OCR, and more than 3000 concept detectors (Section~\ref{sec:concepts}).
The semantic query generation utilized WordNet similarity \cite{Wor}, Point-wise Mutual Information on Wikipedia, and word2vec \cite{Wor} \cite{Tom13} to generate a
mapping that maps the textual event description to the concepts in our vocabulary.
For semantic event search\cite{jiang2015bridging}, our system incorporated various retrieval methods such as Vector Space Model, tf-idf, BM25, language model \cite{CZh01}, etc.
For query by video examples, our system extracted 47 low-level and semantic features, which were all provided to the learning-based search component.
The search components utilized two classifiers: SVM and KRR.
For 100Ex, both SVM and KRR were used.
However, for 010Ex, experiments had shown that only using KRR achieves better performance, 
so the 010Ex runs only utilized prediction results from KRR.
The output of the 47 features from the classifiers were given to Multistage Hybrid Late Fusion to acquire the final fusion results.
More details of the 47 features are in Yu et al.\cite{informedia14}.

Figures~\ref{fig:prespecified} and \ref{fig:adhoc} present the results of our system and other competing systems on the MED14 task.
We can see that our system is significantly better than other competing systems, thus demonstrating the effectiveness of our strategies.
In the following sections, we will detail each of our strategies. However, as the labels for the testing set were never released,
we can only present experimental results on the 20 pre-specified events on the validation set MEDTEST14.

\section{Improvements in Low-Level Features}
\label{sec:feat}
\subsection{Enhancements for Improved Dense Trajectories}
\label{sec:imtraj}

We improve the original Improved Dense Trajectory \cite{Hen13} in two ways. First, temporal scale-invariance is achieved by extracting features under different video playback speeds, which are generated by skipping frames at certain intervals. We denote this new way of feature extraction as Multi-skIp Feature Stacking (MIFS) \cite{Lan141}.  Different from what has been described in Lan et al.\cite{Lan141}, we use the combination of level 0, 2 and 5 to balance speed and performance. 

Second, we propose a new space-time encoding method, dubbed Space-Time Extended Descriptors (STED), that attaches spatial $(x, y)$ and temporal $(t)$ location information to the raw features after PCA-projection \cite{Lan141}. 

As illustrated in Table~\ref{tab:mifsmed}, by using MIFS, we improve MAP of both 100Ex and 010Ex on MEDTEST14 by about 2\%, absolute. We further add STED to the results of MIFS and compared it with Spatial Pyramid Matching (SPM) \cite{Laz06}, a classical space-time encoding method. As can be seen, STED can get similar or better results compared to the results of only using MIFS. SPM can also improve the baseline results, but due to its high dimensionality, it needs large space for storing the resulting feature vectors and is computationally expensive to run the classifiers, thus STED is a more space efficient alternative to incorporate spatial and temporal information into a feature. For details, please see Lan et al.\cite{Lan141}.

\subsection{Features from ImageNet DCNN Models}
\label{sec:koala}
\begin{figure}[tp]
\begin{center}
\begin{tabular}{ccc}
\includegraphics[width=8.0cm,angle=0]{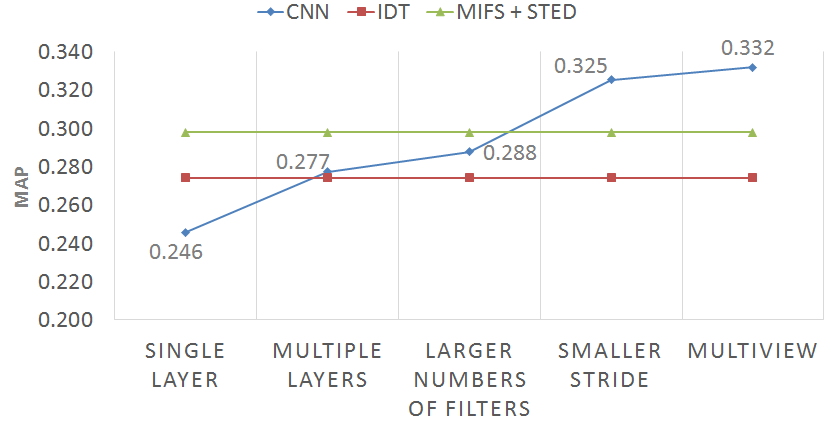} \\
\end{tabular}
\end{center}
\caption{
Performance gains on MEDTEST14 100Ex after CNN structure modifications.
}
\label{fig:cnn}
\end{figure}

In order to leverage the powerful deep learning models in MED, we improved existing DCNN models in two directions: 1) by utilizing more data and 2) by modifying the network structure.
In total, we have extracted a total of 15 different Deep Convolutional Neural Network (DCNN) features in our MED14 system. All models were trained on different subsets of ImageNet. 

We utilized more data by training 3 models\cite{Lan14} on the whole ImageNet dataset consisting of around 14 million labeled images 
and 28,000 classes. 
We took the networks at epoch 5, 6 and 7 and generated features for MED keyframes using the first fully connected layer and probability layer. To generate video features from keyframe-level features, we used both maximum pooling and average pooling for the probability layer and only average pooling for the fully connected layer. This procedure results in 9 DCNN-ImageNet representations for each video. 

To explore the performance of deep models under varying network structures, another 5 models were trained
on the standard ILSVRC 2012 dataset \cite{Den09} which had around 1.28 million images belonging to 1,000 classes.
Two models were trained with six convolutional layers, two models were trained with smaller filters, and one was trained with a larger number of filters. A multi-view representation was used for one of the models. The network structure is as described in Zeiler et al.\cite{Zei14}. Except for different structures among models, the models with the same structures differ in initialization. 
The training process was tuned on the ImageNet ILSVRC 2012 validation set with 50 thousand images.
These models result in another 6 different feature representations. More details and also some further improvements after the MED14 evaluation are described in \cite{Zho14}.

Figure~\ref{fig:cnn} illustrates the improvements on different network structures we have explored within the ILSVRC 2012 training setting. We started with the standard AlexNet but with 6 convolutional layers and the features were computed from the 1,000 dimensional probability layer. Intuitively, the probability output provides a semantic feature representation for each video, where each dimension corresponds to a specific object. 
We can regard this feature as Bag-of-Words with a vocabulary of 1,000 visual objects. The features were then fed into a $\chi^2$-exponential SVM for classification. We only achieved 0.246 MAP on MEDTEST14 100Ex, which is far below IDT. We then explored features from other layers, e.g., $\text{pool}_5$, $\text{fc}_6$ and $\text{fc}_7$. 
Adding multiple layers into the video representation increases the MAP to 0.277. 

We further explored a wider network by doubling the number of filters in each convolutional layer. For example, the standard AlexNet had 256 filters in the 5-th convolutional layer, while we explored the 5-th convolutional layer with 512 filters. This way, the network learns more complex patterns in the images and improved the MAP to 0.288. Following Zeiler et al.\cite{Zei14}, we made the filter size of the first convolutional layer smaller, i.e., reducing it from 11 to 4, and decreased the stride of this layer from 4 to 2. Though this dramatically increased training time for the network due to much more time-consuming convolutional operations on the first convolutional layer, the smaller filter size and stride helped the network capture finer patterns and boosted performance to 0.325, which outperforms the previous versions significantly. 
In the stages described above, we only utilized a single crop from the central 224-by-224 pixels of the video frames, which may lose some helpful visual information. Therefore, we generated a multi-view DCNN feature by producing 10 crops per input frame, which included the top-left, top-right, bottom-left, bottom-right and center crops along with their corresponding mirrored crops. 
The features obtained from the 10 views are subsequently averaged together to acquire a single vector representation.
This further improved performance to 0.332.
%After obtaining the 10 views for the video representation, we predict the event detection results and then apply simple average late fusion over 10 views to get the final prediction. 
The whole exploration of utilizing features extracted from ImageNet pretrained models with different structures raised performance from 0.246 to 0.332, which is a big improvement over state-of-the-art hand-crafted features.

\section{Bridging the Text/Video Semantic Gap}
\label{sec:gap}
%\section{Large-scale Shot-based Semantic Concept}
\label{sec:concepts}
Our shot-based semantic concepts were directly trained on video shots and not still images for the following two reasons: 1) shot-based concepts have minimal domain difference; 2) this allows for action detection. We have found that detectors trained on still images usually do not work well on video, which may suggest that the domain difference between static images and video data such as MED data is significant.  

The shot-based semantic concept detectors were trained with our pipeline based on our previous study on CascadeSVM and a new study on self-paced learning \cite{nips2014} \cite{Kum10}. Our system included more than 3,000 shot-based concept detectors which were trained on around 2.7 million shots using the standard improved dense trajectory features\cite{Hen13}. The detectors are generic and include people, scenes, activities, sports, and fine-grained actions described in \cite{Sho14}. The detectors were trained on several datasets including Semantic Indexing \cite{TRECVID14}, YFCC100M \cite{Yah} and Google Sports \cite{Spo}. YFCC100M and Google Sports are weakly-labeled datasets, i.e. the labels for each video were inferred from the metadata of the videos and not annotated by a human. The notable increase in quantity and quality of our detectors significantly contributed to the improvement in the text-to-video system performance.

Training large-scale concept detectors on big data is very challenging,
thus requiring research on both theoretical and practical perspectives. 
Regarding theoretical progress, we adapted self-paced learning theory, which provided theoretical justification for concept training. Self-paced learning is inspired by the learning process of humans and animals \cite{Kum10,Ben09}, where samples were not learned randomly but organized in a meaningful order: from easier samples to gradually more complex ones. We advanced the theory in two directions: augmenting the learning schemes \cite{LuJ141} and learning from easy and diverse samples \cite{nips2014}. The two studies offer a theoretical foundation for our detector training system. 

As for practical progress, we optimized our pipeline for high-dimensional features (around 100K dimensional dense vector). Specifically, we utilize large shared-memory machines to store the kernel matrices, e.g. 512GB in size, in memory to achieve 8 times speedup in training. This enabled us to efficiently train more than 3,000 concept detectors over 2.7 million shots by self-paced learning \cite{nips2014}. We use around 768 cores in Pittsburgh Supercomputing Center for about 5 weeks, which could be roughly broken down into two parts: low-level feature extraction for 3 weeks and concept training for 2 weeks. For testing, we converted our models to linear models to achieve around 1,000 times speedup in prediction.

In summary, our theoretical and practical progress provided the foundation for developing critical tools for large-scale concepts training on big data. For instance, if we had 500 concepts over 0.5 million shots, then, optimistically speaking, we can finish training within 48 hours on 512 cores, including the raw feature extraction. After getting the models, the prediction for a shot/video only takes 0.125s on a single core with 16GB memory.

\section{Improvements in Indexing/Retrieval}
\label{sec:indexing}
\subsection{Efficient Learning-based Search}
\label{sec:pq}

The most natural way for a human to utilize a system is through an interactive process.
Therefore, to strive for interactive MED, we targeted completing learning-based search over 200,000 videos in 15 minutes on a single machine. 
This is a big challenge for the query by video example pipeline, as we utilized 47 features and around 100 classifiers (SVM \& KRR) to create the final ranked list. The text search pipeline is a lot simpler thus timing is not a big issue. Therefore, we will focus on the query by video example system in the remaining section. To speed up, we performed optimizations in three different directions: 1) decreasing computational requirements, 2) decreasing I/O requirements and 3) utilizing GPUs. Computational requirements were decreased by replacing kernel classifiers with linear classifiers. I/O requirements were decreased by compressing features vectors with Product Quantization\cite{Jeg11} (PQ). GPUs were utilized for fast linear regression and prediction.

\subsubsection{Replacing Kernel Classifiers by Linear Classifiers}
Kernel classifiers are slow during prediction time because to perform prediction on a testing video vector, it is often required to compute the dot-product between the testing video feature and each vector in the training set. For MED14, we had around 5000 training videos, so 5000 dot products were required to predict one video. This is too slow, as preliminary experiments showed that prediction of improved trajectory fisher vectors (IDT-FV, 109056 dimensions) on 200,000 videos required 50 minutes on a NVIDIA K-20 GPU. Therefore, to accelerate this process, we switched to linear classifiers, which requires only one dot product per testing vector, thus in theory we have sped up by 5000x. However, bag-of-word features do not perform well with linear kernels. Therefore, we used the Explicit Feature Map (EFM) \cite{And12} to map all bag-of-word features to a linearly separable space before applying the linear classifier. As the EFM is an approximation, we run the risk of a slight drop in performance. Figures~\ref{fig:pqefm100} and \ref{fig:pqefm10} show the performance difference before and after EFM approximations. For most features, we suffer a slight drop in performance, which is still cost-effective given that prediction speed was sped up by 5000x. 
%Classifier training speed is also improved because we only need to search over one parameter instead of two during cross-validation when using linear classifiers. 
%We saw a 15x speed up for SVM training and a 5x speed up for KRR training. 
%For linear features, the KRR model effectively becomes a linear regression model.

\subsubsection{Feature Compression with Quantization}

\begin{figure}[tp]
\begin{center}
\begin{tabular}{ccc}
\includegraphics[width=8.0cm,angle=0]{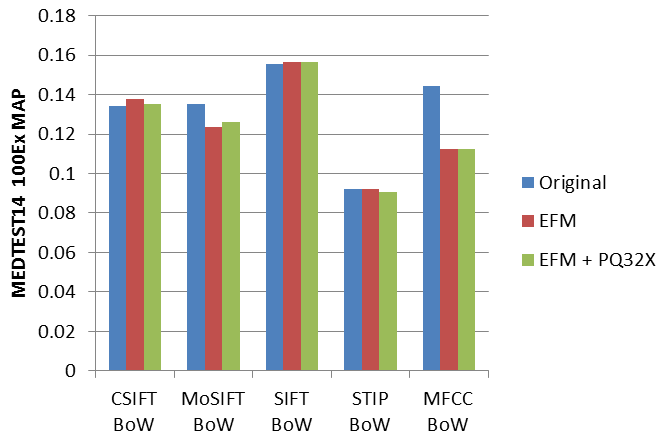} \\
\end{tabular}
\end{center}
\caption{
Performance difference before and after EFM and PQ approximations for MEDTEST14 100Ex.
}
\label{fig:pqefm100}
\end{figure}

\begin{figure}[tp]
\begin{center}
\begin{tabular}{ccc}
\includegraphics[width=8.0cm,angle=0]{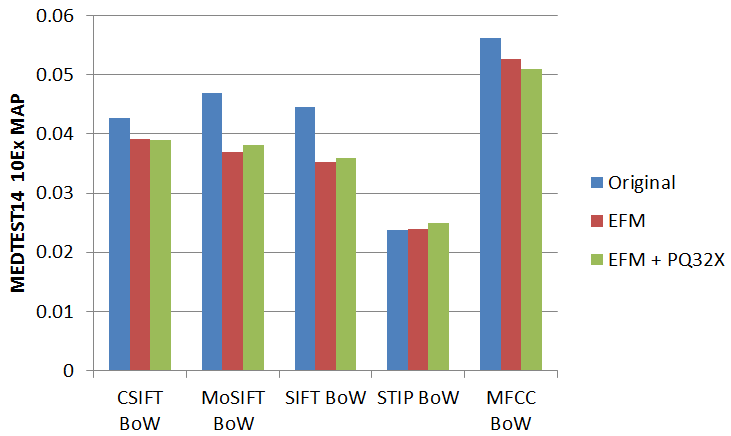} \\
\end{tabular}
\end{center}
\caption{
Performance difference before and after EFM and PQ approximations for MEDTEST14 010Ex.
}
\label{fig:pqefm10}
\end{figure}

\begin{figure}[tp]
\begin{center}
\begin{tabular}{ccc}
\includegraphics[width=8.0cm,angle=0]{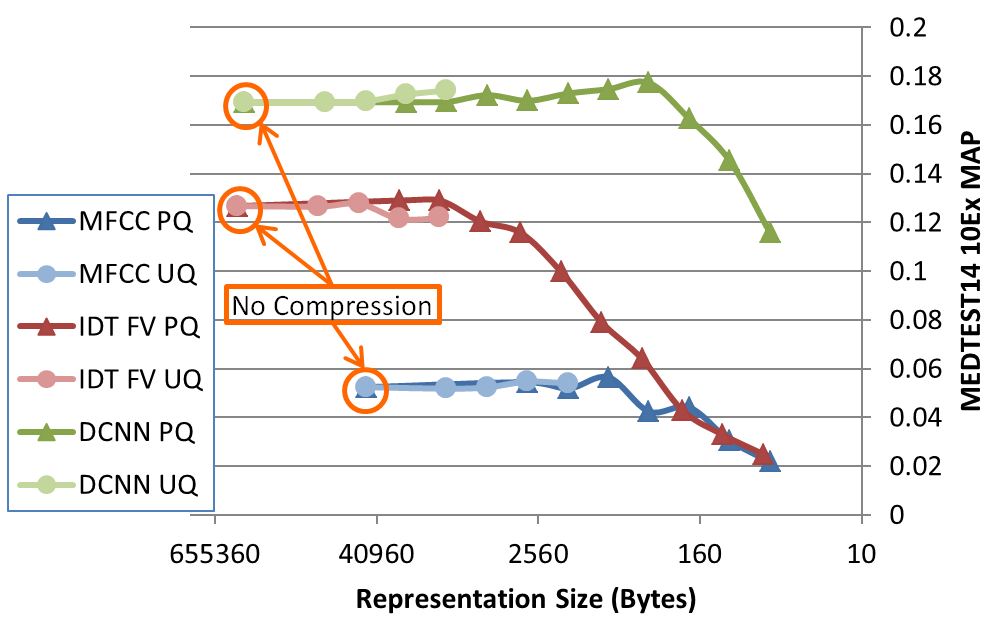} \\
\end{tabular}
\end{center}
\caption{
Performance on MEDTEST14 010Ex under different quantization methods and compression ratios.
}
\label{fig:compression}
\end{figure}

In order to improve I/O performance, we compressed our features using Product Quantization\cite{Jeg11} (PQ). Compression is crucial because reading uncompressed features can take a lot of time. PQ compresses feature vectors by first splitting each feature vector into multiple chunks, and then quantizing each chunk with a 256 word codebook. A 256 word codebook is ideal because cluster assignments can be stored with 1 byte. Therefore, the chunk, which we set to 8 floating point numbers (32 bytes) in our system, is simply represented by 1 byte, thus achieving 32X compression. Also, faster classifier prediction can be done based on the PQ codebooks. More details are in Jegou et al.\cite{Jeg11} and Yu et al.\cite{yu2015content}. However, as PQ performs lossy compression, the quality of the final ranked list may degrade. Figures~\ref{fig:pqefm100} and \ref{fig:pqefm10} shows the performance drop before and after PQ approximation. We can see that there is nearly no performance drop before and after PQ. Figure~\ref{fig:compression} further shows MEDTEST14 010Ex performance when performing quantization under different compression ratios.
We show performance of two quantization methods, PQ and Uniform Quantization (UQ). The basic idea of UQ is to quantize each dimension of all feature vectors into $k$ bins, and each dimension can be represented with $\log_2(k)$ bits.
As we can see, PQ and UQ have similar performance. The problem with UQ is that one can at most achieve 32X compression when $k=2$, but PQ can achieve higher compression ratios by adjusting the size of each chunk.

\subsubsection{Utilizing GPUs for Fast Linear Regression\footnote{For linear features, the KRR model effectively becomes linear regression.} and Linear Classifier Prediction}

Following the TRECVID MED 2014 guidelines, we were limited to a single workstation for learning-based search. Therefore, we utilized all available computing resources on our workstation, which includes CPUs and GPUs. Exploiting the fact that matrix inversion on GPUs are faster than CPUs, we trained our linear regression models on GPUs, which is 4 times faster than running on a 12 core CPU. We also ported the linear classifier prediction step to the GPU, which runs as fast as a 12 core CPU. Our workstation had 2 Intel(R) Xeon(R) CPU E5-2640 6 core processors, 4 NVIDIA TESLA K20's, 128GB RAM, and 10 1T SSDs setup in RAID 10 to increase I/O bandwidth.

\subsubsection{Overall Speed Improvements}

As both EFM and PQ are approximations, we quantified the drop in performance when both methods were used. The results are shown in Table~\ref{tab:efmpq}. We see a 3\% relative drop in performance for 100Ex and a slight gain in performance for 010Ex. Despite slight drop in performance, speed has been substantially decreased. We have sped up our system by 16 times for learning-based search with a cost of 3\% relative drop in performance, which is negligible given the large efficiency gain.

\begin{table}[tp]
\centering
	\begin{tabular}{ | @{\hspace{0pt}}C{3.0cm}@{\hspace{0pt}} || @{\hspace{0pt}}C{1.2cm}@{\hspace{0pt}} | @{\hspace{0pt}}C{1.2cm}@{\hspace{0pt}} | @{\hspace{0pt}}C{2.6cm}@{\hspace{0pt}} |  }
		\hline
		 \multirow{2}{*}{}	&	\multicolumn{2}{c|}{MEDTEST14 MAP}	&	\multirow{2}{*}{ {\vspace{10pt}}{\hspace{-8pt}} \begin{tabular}{c}100Ex Learning-based \\ Search Timing (s) \end{tabular}} \\ \cline{2-3} 
		 & 100Ex & 010Ex & \\ \hline
	    No EFM, No PQ, with GMM features$^\alpha$ &	0.405 & 0.266 & 17580$^\beta$ \\\hline
		EFM, PQ, no GMM features & 0.394$^\gamma$ & 0.270 & 1068 \\ \hline
		Relative Improvement & -2.7\% & 1.5\% & 1646\% \\ \hline
	\end{tabular} 
\caption{Performance of different features and fusion methods.}
\label{tab:efmpq}
\end{table}

\blfootnote{$\alpha\,\,$: Used in our MED13 system\cite{Lan13}.}
\blfootnote{$\beta\,\,$: Extrapolated timing for our MED13 system\cite{Lan13}.}
\blfootnote{$\gamma\,\,$: A modified MHLF was used so that it is compatible with features of the MED13 system, thus leading to slightly different numbers than Table~\ref{tab:res14}.}

\subsection{Multistage Hybrid Late Fusion Method}
\label{sec:ming}

\begin{figure}[tp]
\begin{center}
\begin{tabular}{ccc}
\includegraphics[width=8.0cm,angle=0]{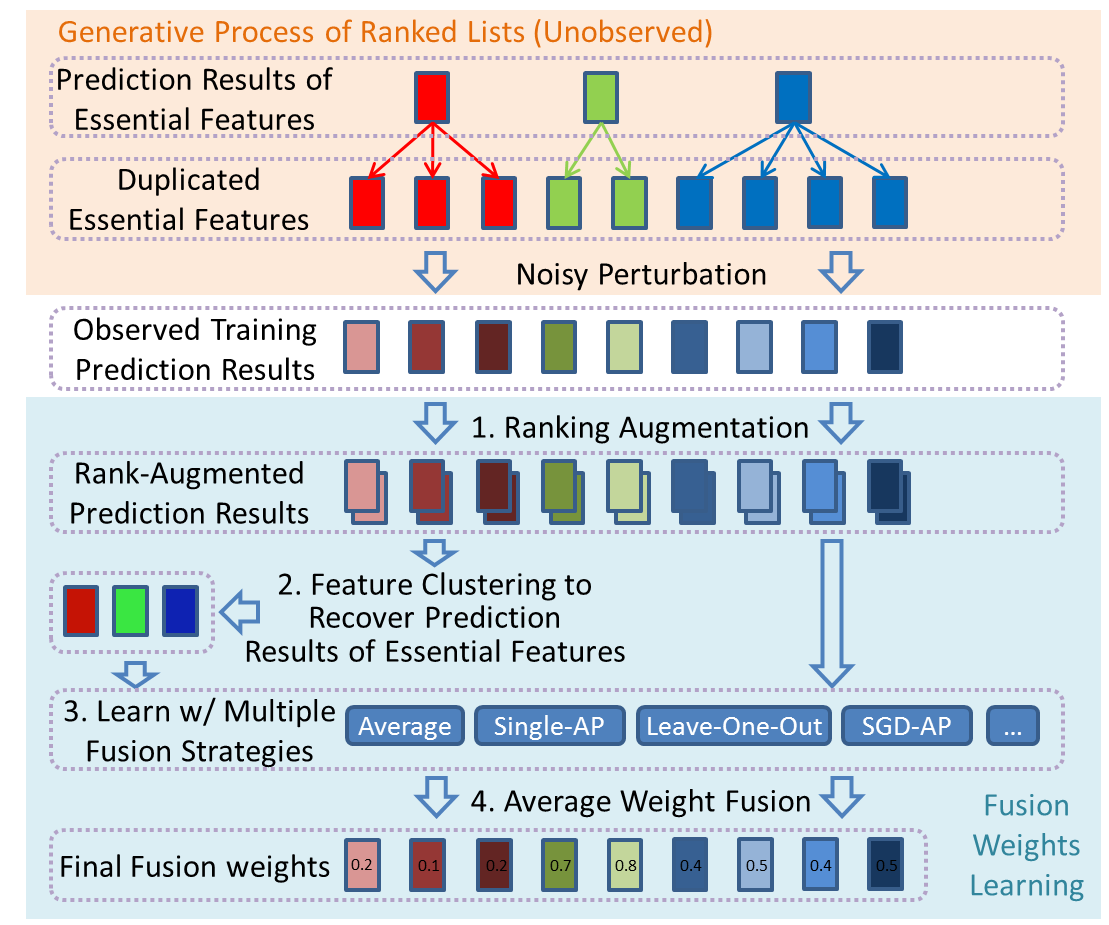} \\
\end{tabular}
\end{center}
\caption{
Intuition and pipeline of Multistage Hybrid Late Fusion.
}
\label{fig:fusion}
\end{figure}

For a given query, the goal of fusion is to learn the weights of different modalities according to the effectiveness of each feature.
A simple way to learn modality specific weights can be by training a linear regression model on held-out data from the training set.
However, this approach is usually not as stable as if the held-out set is small, the learned weights tend to overfit.
To this end, we propose a new learning based late fusion algorithm, named the ``Multistage Hybrid Late Fusion'' (MHLF) as shown in Figure~\ref{fig:fusion}. 
The MHLF is designed based on the following three key observations:\\
\noindent\textbf{1. Ranking information is not explicitly modeled in the prediction scores.} Therefore, step 1 in Figure~\ref{fig:fusion}
augments the original prediction scores with ranking information. \\
\noindent\textbf{2. Prediction scores from different features contain duplicate information and should not be  na\"{\i}vely averaged.}
Duplicate
information comes from different features using the same basic feature. For example, SIFT-BoW and CSIFT-FV are all SIFT-based
and their ranked lists are usually highly correlated.
We propose to model such highly correlated ranked-lists as a generative process.
The assumption is that there are many ``essential features'', whose classifiers generate noise free ranked lists.
However, these essential features goes through a duplication and noisy perturbation process, thus
what we observe are noisy prediction results.
Therefore, to recover the essential features, we perform PCA-Tree clustering as shown in step 2 of Figure~\ref{fig:fusion}.
The cluster centers corresponds to a ``cleaner'' version of the prediction results and can be viewed as an estimate of an essential feature.
These recovered essential features, and also the original prediction scores are all provided to the next hybrid fusion step. \\
\noindent\textbf{3. Prediction scores contain random noise and directly learning fusion weights on top may lead to overfitting.}
To deal with this issue, MHLF utilizes hybrid strategies to acquire a more robust fusion weight estimate.
The intuition is that each fusion strategy can be viewed as a random observation of a ``ground-truth fusion strategy''.
Since there is no single fusion strategy that performs better than others on all queries,
sampling multiple strategies and averaging them is a simple and effective method
to acquire a more stable estimate of fusion weights. 
The key fusion strategies include:
\begin{enumerate}
\item Average fusion: each feature gets equal weight.
\item Single-AP: the weights of each feature is its average precision (AP) on the held-out set.
\item Leave-One-Out: the weights of a feature is the AP performance drop when removing this feature from an average fusion run.
\item SGD-AP: performs stochastic gradient descent which maximizes average precision as the loss function.
\end{enumerate}

\begin{figure}[tp]
\begin{center}
\begin{tabular}{ccc}
\includegraphics[width=6.0cm,angle=0]{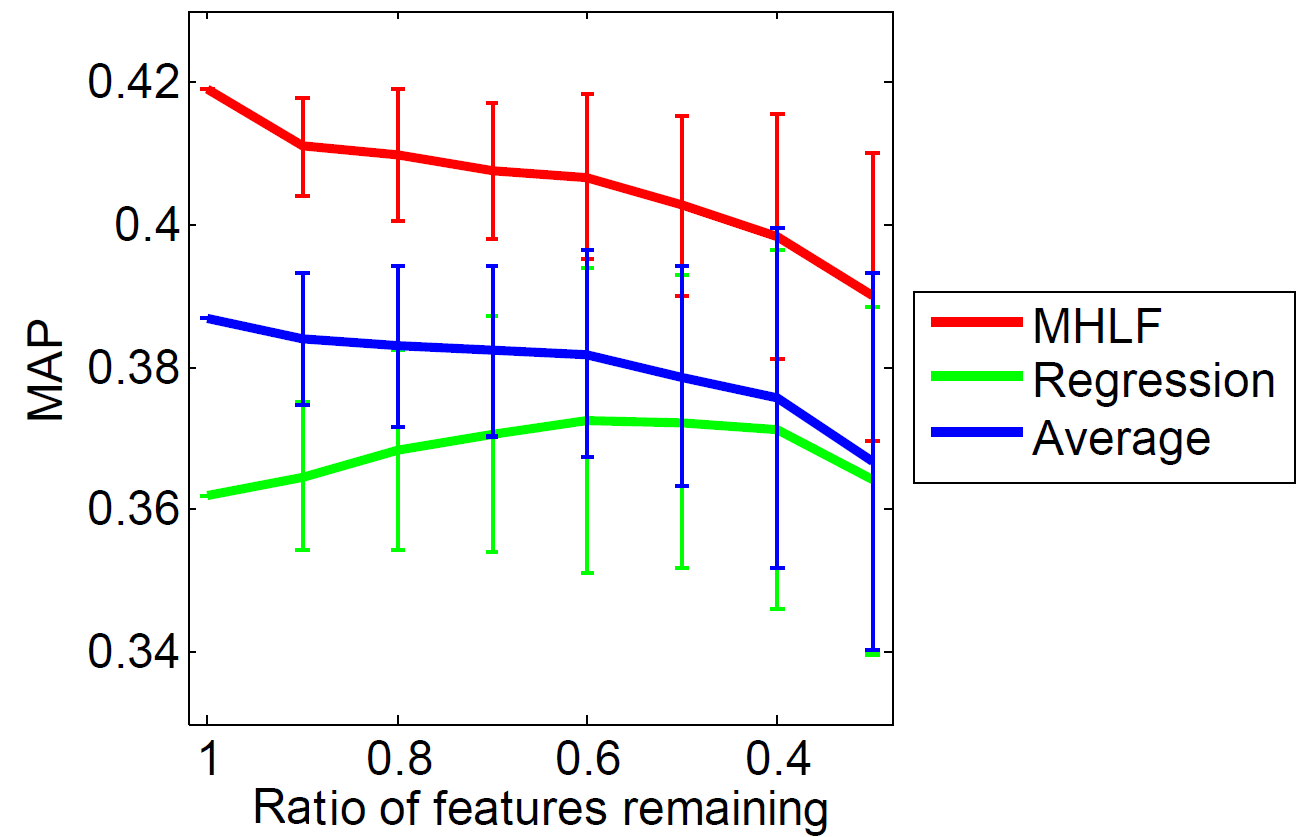} \\
\end{tabular}
\end{center}
\caption{
Results of fusion based on randomly sampled subsets of features, where the number of randomly sampled features varies from all features to 30\% of the features.
The sampling was repeated 60 times, and the 95\% confidence intervals are shown.
}
\label{fig:robust}
\end{figure}

Results on the key features of MEDTEST14 and final fusion results are shown in Table~\ref{tab:res14}.
All these results were based on prediction scores from 32X PQ.
As we can see, MHLF is superior than average fusion and linear regression fusion.
We also performed robustness tests on our fusion algorithm as shown in Figure~\ref{fig:robust}.
In this experiment, we randomly removed a subset of features from the original set of 47 features and ran the different fusion methods. 
As we can see, as we gradually remove features, MHLF is still consistently better than the other two baseline fusion algorithms,
thus demonstrating the robustness of MHLF.

%(2) [Section 6.2] What are the pros and cons of methods in Table 3 (res14)? If these methods are complementary, which one is the most effective for which types of events?

From Table~\ref{tab:res14}, we can also compare the relative performance of single features and encoding methods.
For static image features, it is clear that multi-view DCNN outperforms handcrafted features such as SIFT and CSIFT.
For motion features, MIFS + STED is significantly better than MoSIFT and STIP.
Encoding wise, we can see that Fisher vector (FV) encodings are in general better than BoW encodings.
%Furthermore, IDT-based and CNN-based features are significantly better than the traditional SIFT, CSIFT, MoSIFT features.
Also, in general KRR performs better than SVM in the 010Ex scenario, so for the 010Ex fusion runs we only used prediction results from KRR.
Finally, if there were resource constraints in feature extraction, 
combining the 3 core features: MFCC BoW, MIFS + STED, and multi-view DCNN can 
achieve around 90\% of the full system's performance.
Among these 3 core features, MFCC excels on events such as ``Tuning musical instrument'' and ``Town hall meeting'', where audio
such as instrument sounds or speech is an important cue.
MIFS + STED performs well on events such as ``Rock climbing'' and ``Winning a race without a vehicle'', where the action of people
is crucial in determining if a video is relevant.
Finally, multi-view DCNN achieves high performance on events which have discriminative objects,
such as honeycomb for the ``Beekeeping'' event, and cars in the ``Parking a vehicle'' event.

\begin{table}[tp]
\centering
	\begin{tabular}{ | @{\hspace{0pt}}C{2.6cm}@{\hspace{0pt}} || @{\hspace{0pt}}C{1.2cm}@{\hspace{0pt}} | @{\hspace{0pt}}C{1.2cm}@{\hspace{0pt}} | @{\hspace{0pt}}C{1.2cm}@{\hspace{0pt}} | @{\hspace{0pt}}C{1.2cm}@{\hspace{0pt}} | }
		\hline
		 
Condition	&	\multicolumn{2}{c|}{10Ex}	&	\multicolumn{2}{c|}{100Ex} \\ \hline
Classifier	&	KRR	&	SVM	&	KRR	&	SVM \\ \hline\hline
STIP SpBoW EFM	&	0.026	&	0.025	&	0.087	&	0.091 \\ \hline
SIFT SpBoW EFM	&	0.042	&	0.036	&	0.145	&	0.157 \\ \hline
MoSIFT SpBoW EFM	&	0.045	&	0.038	&	0.110	&	0.126 \\ \hline
CSIFT SpBoW EFM	&	0.046	&	0.039	&	0.143	&	0.135 \\ \hline
MFCC BoW EFM	&	0.057	&	0.051	&	0.101	&	0.112 \\ \hline
CSIFT FV	&	0.065	&	0.051	&	0.157	&	0.140 \\ \hline
SIFT FV	&	0.066	&	0.060	&	0.162	&	0.157 \\ \hline
STIP FV	&	0.073	&	0.074	&	0.140	&	0.140 \\ \hline
MoSIFT FV	&	0.081	&	0.083	&	0.179	&	0.184 \\ \hline
IDT FV	&	0.135	&	0.128	&	0.270	&	0.268 \\ \hline
MIFS + STED	&	0.161	&	0.142	&	0.292	&	0.277 \\ \hline
Multi-view DCNN	&	0.187	&	0.167	&	0.319	&	0.299 \\ \hline
MHLF, MIFS + STED \& multi-view DCNN &	\multicolumn{2}{c|}{0.215}	&	\multicolumn{2}{c|}{0.353} \\ \hline
MHLF, MIFS + STED \& multi-view DCNN \& MFCC BoW &	\multicolumn{2}{c|}{0.237}	&	\multicolumn{2}{c|}{0.389} \\ \hline
Linear Regression Fusion, 47 Features	&	\multicolumn{2}{c|}{0.250}	&	\multicolumn{2}{c|}{0.362} \\ \hline
Average Fusion, 47 Features	&	\multicolumn{2}{c|}{0.252}	&	\multicolumn{2}{c|}{0.387} \\ \hline
MHLF, 47 Features	&	\multicolumn{2}{c|}{\textbf{0.285}}	&	\multicolumn{2}{c|}{\textbf{0.419}} \\ \hline
	\end{tabular} 
\caption{MAP performance of different features and fusion methods. The testing features have all gone through 32X PQ compression, so the results are slightly lower than the non-approximated results reported in Table~\ref{tab:mifsmed} and Figure~\ref{fig:cnn}.}
\label{tab:res14}
\end{table}

\subsection{Self-Paced Reranking}
\label{sec:reranking}
Our PRF system was implemented according to Self-Paced Reranking (SPaR) detailed in Jiang et al.\cite{LuJ141}. SPaR represents a general method of addressing multimodal pseudo relevance feedback for SQ/000Ex video search. As opposed to utilizing all samples to learn a model simultaneously, the proposed model is learned gradually from easy to more complex samples. In the context of the reranking problem, the easy samples are the top-ranked videos that have smaller loss. As the name ``self-paced'' suggests, in every iteration, SPaR examines the ``easiness'' of each sample based on what it has already learned, and adaptively determines their weights to be used in the subsequent iterations.
The mixture weighting/scheme self-paced function was used, since we empirically found it outperforms the binary self-paced function on the validation set~\cite{jiang2015bridging}. Since the starting values can significantly affect final performance, we used the reasonable starting values generated by MMPRF \cite{MMPRF}. The high-level features used were ASR, OCR, and semantic visual concepts. The low-level features were DCNN, IDT and MFCC features. We did not run PRF for SQ since our 000Ex and SQ runs are very similar. The final results were computed by averaging the initial ranked list with the reranked list.
This is beneficial because for the 000Ex case, the initial ranked list is from semantic search (high-level features), whereas the reranked list is from learning-based search (low-level features), and leveraging high-level and low-level features usually yields better performance\cite{LuJ12}. To be prudent, the number of iterations is no more than 2 in our final submissions. 
%For more details, please refer to \cite{MMPRF} and \cite{LuJ141}.

The contribution of our reranking methods is evident because the reranking method is the only difference between our noPRF runs and PRF runs as shown in Figure~\ref{fig:prespecified} and \ref{fig:adhoc}. According to the MAP on the testing set of MED14, our reranking method boosted the MAP of the 000Ex system by a relative 16.8\% for pre-specified events and a relative 51.2\% for ad-hoc events. Besides, it also boosted the 010Ex system by a relative 4.2\% for pre-specified events, and a relative 13.7\% for ad-hoc events. This observation is consistent with the ones reported in previous work\cite{MMPRF,LuJ141}. Note that the ad-hoc queries are very challenging because the query is unknown to the system beforehand. As we can see, our reranking methods still managed to yield significant improvement on ad-hoc events. More reranking results on MEDTEST14 data can be found in Jiang et al.~\cite{jiang2015bridging}.
%Also, as shown in Yu et al.\cite{yu2015content}, reranking is not only effective, but also can be implemented and ran in an efficient way.

It is interesting that our 000Ex system for ad-hoc events outperforms 010Ex systems from many other teams. In MED14, the difference between the best 000Ex with PRF (17.7\%) and the best 010Ex noPRF (18.2\%) is marginal. In MED13, however, this difference was very large where the best 000Ex and 010Ex system was 10.1\% and 21.2\%\footnote{The runs in different years are not comparable since different queries were used.} respectively. This observation suggests that the gap of real-world 000Ex event search system is shrinking rapidly. 
We attribute the improvement of the 000Ex system to the following key reasons: 1) improved semantic concept detectors (Section~\ref{sec:concepts}), 2) improvement achieved by the reranking algorithm SPaR, and 3) reasonable queries formulated by human experts.

\section{Conclusion and Future Work}
\label{sec:conclusion}
We have described multiple strategies to enhance both the accuracy and speed of content-based video retrieval systems.
Overall, the main conclusions are:
1) IDT-based and CNN-based features are the current best motion and static image feature, 
2) semantic concept detectors trained from big data are effective,
3) EFM and PQ compression can significantly speed up the system with only a negligible drop in accuracy,
4) MHLF fusion, which fuses multiple fusion strategies, is robust, and
5) reranking is an effective way to enhance accuracy.
Looking into the future, we believe that current systems can already achieve reasonable accuracy, but speed is still a big issue.
Efficiently extracting features, indexing and searching the billions of videos online will be the next big challenge.

\section*{Acknowledgments}
This work was partially supported by the US Department of Defense the U. S. Army Research Office (W911NF-13-1-0277),  National Science Foundation under Grant Number IIS-12511827 and the Intelligence Advanced Research Projects Activity (IARPA) via Department of Interior National Business Center contract number D11PC20068. The U.S. government is authorized to reproduce and distribute reprints for Governmental purposes notwithstanding any copyright annotation thereon. Disclaimer: The views and conclusions contained herein are those of the authors and should not be interpreted as necessarily representing the official policies or endorsements, either expressed or implied, of IARPA, DoI/NBC, or the U.S. Government.
This work used the Extreme Science and Engineering Discovery Environment (XSEDE), which is supported by National Science Foundation grant number OCI-1053575. Specifically, it used the Blacklight system at the Pittsburgh Supercomputing Center (PSC).

\bibliographystyle{ieee}
\bibliography{egbib}

%\appendix
%\input{Appendix}

\begin{biography}
\profile{Shoou-I}[]{Yu}[]{
received the B.S. in Computer Science and Information Engineering from National
Taiwan University, Taiwan in 2009. He is now a Ph.D. student in Language Technologies
Institute, Carnegie Mellon University. His research interests include multi-object tracking and multimedia retrieval.
}
\profile{Yi}[]{Yang}[]{
Yi Yang received the PhD degree from Zhejiang University in 2010. He was a postdoc research fellow with the School of Computer Science at Carnegie Mellon University. He is now an Associate Professor with University of Technology Sydney. His research interest include multimedia, computer vision and machine learning.
}
\profile{Zhongwen}[]{Xu}[]{received the B.E. in Computer Science and Technology from Zhejiang University, China in 2013. He is now a Ph.D. student at Centre for Quantum Computation \& Intelligent Systems, University of Technology Sydney. His research interests are on computer vision and deep learning, especially for video analysis.
}
\profile{Shicheng}[]{Xu}[]{
received the B.S. in Computer Science and Engineering from Zhejiang University, Hangzhou, Zhejiang, China, in 2014. He is now a Visiting Researcher in Language Technology Institute, Carnegie Mellon University. His research interests include multimedia analysis and multimedia retrieval.
}
\profile{Deyu}[]{Meng}[]{
received the B.Sc., M.Sc., and Ph.D degrees in 2001, 2004, and 2008, respectively, from Xi’an Jiaotong University, Xi’an, China. He is currently an associate professor with the Institute for Information and System Sciences, School of Mathematics and Statistics, Xi’an Jiaotong University. From August 2012 to July 2014, he was a visiting scholar in Carnegie Mellon University. His current research interests include machine learning, computer vision, multimedia analysis and other related topics.
}
\profile{Zexi}[]{Mao}[]{
Zexi Mao received the B.S. in Computer Science and Technology from Zhejiang University, China in 2013, and the M.S. in Language Technologies from Carnegie Mellon University, USA in 2015. He is currently a Data Engineer at Jetlore, Inc.
}
\profile{Zhigang}[]{Ma}[]{
is now a Postdoctoral Research Fellow with the School of Computer Science, Carnegie Mellon University, Pittsburgh, PA. His research interest is mainly on machine learning and its applications to multimedia analysis and computer vision. He has authored or co-authored more than 20 scientific articles at top venues, including the IEEE T-PAMI, T-MM, IJCV, ACM MM, CVPR, AAAI and IJCAI. He was a PC member for ACM MM 2014; a TPC member for ICME 2014 and 2015; a TPC member for ICMR 2015 and a PC member for IJCAI 2015. He is also an invited reviewer for IEEE Transactions on Multimedia, IEEE Transactions on Cybernetics, Multimedia Tools and Applications, Neurocomputing, Computer Vision and Image Understanding. Dr. Ma received the Outstanding PhD thesis award from SIGMM and the best PhD thesis award from Gruppo Italiano Ricercatori in Pattern Recognition, Italy. 
}
\profile{Ming}[]{Lin}[]{
Ming Lin received his Bachelor  and Doctor degree in the Department of Automation from Tsinghua University, Beijing, China, in 2008 and 2014. He is a Postdoctoral Research Fellow in the School of Computer Science at Carnegie Mellon University. His research interest is mainly on machine learning theory and its applications in computer vision.
}
\profile{Xuanchong}[]{Li}[]{
Xuanchong Li received B.E. in computer science and technology from Zhejiang University, China in 2012. He is now a master student in Carnegie Mellon University. His research interest includes computer vision, machine learning.
}
\profile{Huan}[]{Li}[]{
Huan received the PhD degree of computer science from Beihang University, China in 2012. She is currently a software engineer at Microsoft. Her research interest is mainly on machine learning and its applications to multimedia analysis and computer vision.
}
\profile{Zhenzhong}[]{Lan}[]{
received the B.S. in software engineering and statistics from Sun Yat-sen University,
China in 2010. He is now a Ph.D. student at Language Technologies Institute, Carnegie Mellon
University. His research interests include computer vision and multimedia retrieval.
}
\profile{Lu}[]{Jiang}[]{
Lu Jiang received his M.Sc. degree in Computer Science in 2011 and B.Sc. degree in Software Engineering in 2008, both from Xi’an Jiaotong University. Currently, he is a Ph.D candidate at school of computer science, Carnegie Mellon University. His research is focused on multimedia, machine learning, and big data.
}
\profile{Alexander G.}[]{Hauptmann}[]{
received the B.A. and M.A. degrees in psychology from The Johns Hopkins University, Baltimore,
MD, USA, in 1982, the ``Diplom'' in computer science from the Technische Universit\"{a}t Berlin,Berlin, Germany, in 1984,
and the Ph.D. degree in computer science from Carnegie Mellon University (CMU), Pittsburgh, PA, USA in 1991. He is a Principal
Systems'' Scientist in the CMU Computer Science Department and also a faculty member 
with CMU's Language Technologies Institute. His research combines
the areas of multimedia analysis and retrieval, man-machine interfaces,
language processing, and machine learning. He is currently leading the
Informedia project which engages in understanding of video data ranging
from news to surveillance, Internet video for applications in general
retrieval as well as healthcare.
}
\profile{Chuang}[]{Gan}[]{
Chuang Gan received the B.E. in Electronic Engineering from Beihang University, China in 2013. He is now a Ph.D. student at Institute for Interdisciplinary Information Sciences, Tsinghua University. His research interests are on computer vision and machine learning, especially for large-scale video analysis.
}
\profile{Xingzhong}[]{Du}[]{
received the B.S. in Software Engineering and M.S. in Computer Science from Nanjing University, China in 2010 and 2013 respectively. He is now a Ph.D. student at the University of Queensland. His research interests are on content-based video recommender system, video database and surveillance event detection.
}
\profile{Xiaojun}[]{Chang}[]{
is a Ph.D. student at University of Technology Sydney. His research interests include machine learning, data mining and computer vision.
His publications appear in proceedings of prestigious international conference like ICML, ACM MM, AAAI, IJCAI and etc.
}

\end{biography}

\end{document}